\documentclass[%
 reprint,
 amsmath,amssymb,
]{revtex4}
\usepackage[section]{placeins}
\let\Oldsection\section
\renewcommand{\section}{\FloatBarrier\Oldsection}

\let\Oldsubsection\subsection
\renewcommand{\subsection}{\FloatBarrier\Oldsubsection}

\let\Oldsubsubsection\subsubsection
\renewcommand{\subsubsection}{\FloatBarrier\Oldsubsubsection}
\usepackage{graphicx}
\usepackage{dcolumn}
\usepackage{bm}

\begin{document}

\preprint{APS/123-QED}
\title{Optical Flip-Flops and Shift Registers from populations and coherences in multi-level systems using stimulated Raman adiabatic passage}

\author{Dawit Hiluf}
\email{dawit@post.bgu.ac.il}
\affiliation{Ben-Gurion University of Negev department of Physical Chemistry, Be'er-Sheva 84105, Israel}
\affiliation{Physics Department, Mekelle University, P.O.Box 231, Mekelle, Ethiopia.}
\author{Yonatan Dubi}
\email{jdubi@bgu.ac.il}
\affiliation{Ben-Gurion University of Negev department of Physical Chemistry, Be'er-Sheva 84105, Israel}
\affiliation{Ilse-Katz Institute for Nanoscale Science and Technology, Ben-Gurion University of the Negev, Beer-Sheva 84105, Israel}
\date{\today}
\begin{abstract}
In digital circuits, a Flip-Flop (FF)  is a circuit element that has two stable states which can be used to store and remember state information.  The state of the circuit can be changed by applying signals to the control input.  FFs are the basic building blocks of sequential logic circuits, as logic gates AND,OR, NOT are the basic building block for combinational logic circuits, and are therefore necessary for any computations involving memory.  Consequently, the design and implementation of FFs can be considered as a pre-requisite for memory machine design. Here we present the design of an optical FFs in an atomic multi-level system, based on the optical manipulation of populations and coherences using stimulated Raman adiabatic passage. We first demonstrate that both populations and coherences can be transferred over multistate systems. We then propose the design of toggle-FFs, Delay-FFs, and Serial-in Serial-Out (SISO) shift registers using such systems. For the design of the filp-flops we use a three level $\Lambda$-type system. In order to design SISO shift registers we concatenate two $\Lambda$-type systems and construct an "M"-type scheme, and similarly concatenating three $\Lambda$-type system we are able to obtain a seven level system. By concatenating we are able to use output of one three level $\Lambda$-type system serve as input of another three level $\Lambda$-type system.  On top of using populations for design of logic gates, we uniquely exploit the coherences for logic machine, which provides an additional degree of freedom which can be used for the design of computing elements. 
\begin{description}
\item[Usage]
Secondary publications and information retrieval purposes.
\item[PACS numbers]
May be entered using the \verb+\pacs{#1}+ command.
\item[Structure]
You may use the \texttt{description} environment to structure your abstract;
use the optional argument of the \verb+\item+ command to give the category of each item. 
\end{description}
\end{abstract}

\pacs{Valid PACS appear here}
\maketitle


\section{Introduction}
The way towards miniaturization and reduced power consumption of computing machines follows Moore's law, the observation that the number of transistors on an integrated circuit doubles about every two years \cite{moore2010cramming}. Since the processing speed of a computer increases as the distance between transistors on  integrated circuit decreases, Moore's law explained how computers can simultaneously drop in price while improving the performance. The correlation provided by Moore's law  sets the standard benchmark for evaluating the progress of the processor industry. Based on the trend so far it is expected that as the reduction in size of  transistors continues, eventually the realm of atomic length-scales will be reached. Consequently, a paradigm shift in the way we manufacture and perform computing will be needed. In the bottom-up approach, advocated by Feynman in his famous lecture, \emph{there is plenty of room at the bottom} \cite{feynman1960there}, one can manipulate the atoms/molecules individually to build the desired machine. This approach spurred immense research effort to define, construct and manipulate new computing elements, including Quantum Computing, Neural Networks based computing, DNA computing, Quantum Cellular Automata, Single Electron Transistor (SET) and more \cite{Calude2000,remacle2008inter,Lippmann1987,Berman1998}.

An interesting approach is to use molecular systems as processors that can perform logic computations \cite{Andreasson2004,DeSilva2005,PrasannadeSilva2000,Remacle:2001ab}. Within this scheme, molecules are excited optically, and the optical excitation serves both for reading and writing and gate operation. In a series of papers \cite{remacle2008inter,remacle2006all,remaclepar,Remacle:2001aa}, Remacle and Levine suggested using the mechanism of stimulated Raman Adiabatic Passage (STIRAP) protocol to perform computations. With STIRAP one can coherently transfer populations and generate coherences between the quantum states of the molecular system \cite{Gaubatz1990} (see recent reviews in \cite{shore98, Vitanov2016,Bergmann2015}), which can be used to show that a three-level lambda system can perform full addition and subtraction \cite{remacle2006all} and logic operations  \cite{remacle2006all,beil2011logic,2013arXiv1306.2132G,Remacle:2001aa}. Beyond binary states, it was demonstrated how one could perform multivalued logic using electrical excitation \cite{klein2007transcending, klein2010ternary} . 

To proceed in making molecular-based computing machines, one has to go beyond logic elements. In this paper we propose a design for an all-optical serial-in serial-out (SISO) shift register. Shift registers perform two basic functions, namely the storage and transfer of data in binary form, and are essentially finite-state machines \cite{kohavi2010switching}, required for the construction of digital circuits. We introduce different types of STIRAP-based flip-flops (FF), the basic building blocks of sequential logic circuits, from which the shift register is constructed. Notably, we demonstrate the use of both populations and quantum coherences in designing the shift registers, using two forms of STIRAP. The use of coherences as a resource for classical computing operations 
enables the construction of the shift register from a minimal system, in the spirit of maximal size reduction of the circuit elements. 


The outline of the paper is as follows. In section (\ref{sec:thesystem}) we introduce the system and model. In section (\ref{sec:results}) we present the STIRAP dynamics of the system. In section (\ref{sec:logicmachine}) we discuss how we design the FF and SISO shift registers using the STIRAP mechanism.  Section (\ref{sec:conclusion}) is devoted to a summary and outlook.

\section{System and Model}
\label{sec:thesystem}
\begin{figure}[h!]
\begin{center}
\includegraphics[width=4.0 in]{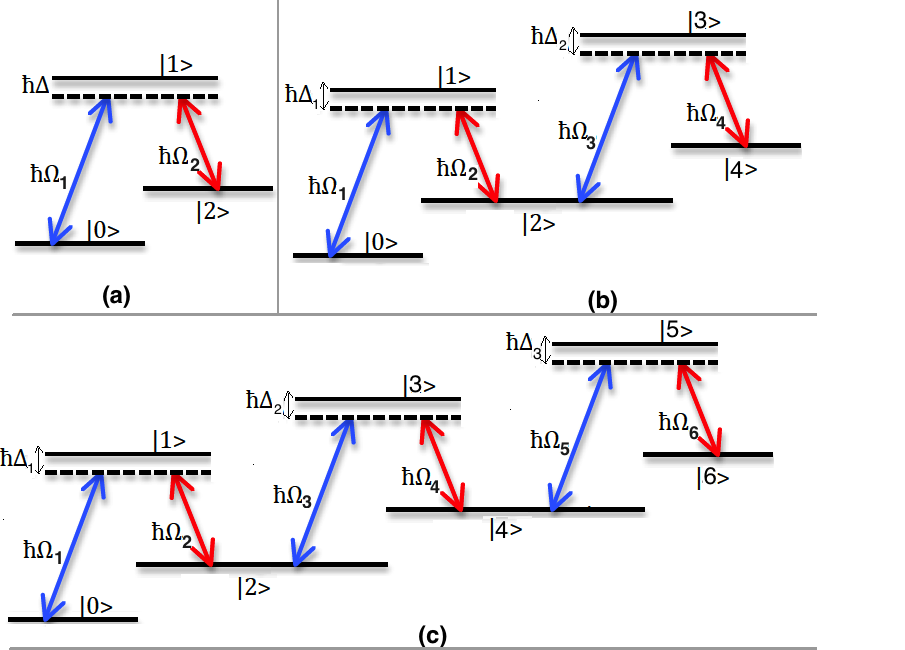}
\caption{(a) Three level system coupled by pump and Stokes laser along with detuning $\hbar\Delta$  (b) Five level system, constructed out of two communicating three level $\Lambda$-type systems. We consider single detuning $\hbar\Delta_1=\hbar\Delta_2=\hbar\Delta$ (c) Seven level system, constructed out of three communicating three level $\Lambda$-type systems. We consider single detuning $\hbar\Delta_1=\hbar\Delta_2=\hbar\Delta_3=\hbar\Delta$. The pump and Stokes pulses are plotted in blue and red lines, respectively.}
\label{MLevel}
\end{center}
\end{figure}

For constructing an all-optical molecular logic, the minimal setting is a so-called  $\Lambda$-system as  shown in Fig.~(\ref{MLevel}a) with states $|0\rangle$, $|1\rangle$ and $|2\rangle$ \cite{remacle2008inter,Remacle:2001aa,Remacle:2001ab}.  The levels $|0\rangle\leftrightarrow |1\rangle$ and $|1\rangle\leftrightarrow |2\rangle$ are coupled by laser fields $\Omega_p \left(t\right)$ and $\Omega_s \left(t\right)$ respectively. The laser is off by a detuning energy $\Delta$, which is the difference between the laser frequency and the Bohr frequency (the energy level difference). The pulse durations used are much shorter compared to relaxation times in the system. We consider an adiabatic population transfer, where a strong pulse is able to transfer all the population, from $|0\rangle$ to $|2\rangle$ \cite{Bergmann2001}. 
To construct an all-optical shift register, the $\Lambda$-system needs to be extended to an M-type system , which extends the system to $N$ levels ($N$ is odd) (see examples for $N=5$ and $N=7$, discussed in the text, in Fig.~(\ref{MLevel}b-c)). A possible physical realization for such systems are Rare-Earth-metal ions doped in solids, exhibiting sharp optical transitions and long decoherence times, which in turn facilitate the implementation of coherent adiabatic processes. A demonstrated example is Pr$^{3+}$:Y$_2$SiO$_5$ (Pr:YSO) \cite{klein2008experimental, beil2008electromagnetically, Klein:2007aa,beil2011logic}, that has three nearly degenerate ground states and three nearly degenerate excited states, with hyperfine states spaced between 4.6 MHz and 17.3 MHz from each other. All the three ground states can be coupled to all three excited states, allowing for nine possible transitions. Out of which we focus on the coupling schemes shown in Fig.(\ref{MLevel}) for three level, five level, and seven level systems.

The optical transitions in Pr:YSO exhibit very narrow homogenous line widths of the order of a few kHz. However, due to crystal impurities, different ions see slightly different crystal fields and the optical line is inhomogeneously  broadened by several GHz. Due to this inhomogeneous broadening, a laser field with single frequency will always address several ions in the inhomogeneously broadened line. 
 \begin{figure}[h!]
\begin{center}
\includegraphics[width=3.0 in]{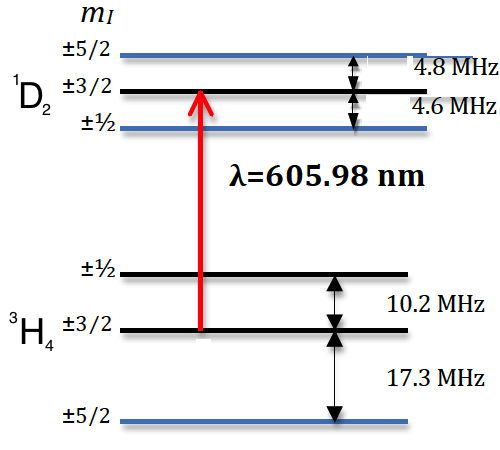}
\caption{(Color online) The energy level of Pr$^{3+}$:Y$_2$SiO$_5$ (Pr:YSO). All the three ground states can be coupled to all three excited states, allowing for nine possible transitions.}
\label{EnergyLevel}
\end{center}
\end{figure}

Because it is possible to have nine transitions by coupling the three ground states to all three excited states shown in the energy diagram, Fig.~(\ref{EnergyLevel}), of Pr:YSO and because the optical line is inhomogeneously broadened  a  laser pulse with single frequency will address several ions and hence by making use of  laser field with different values one is able to have three types of ions with $\Lambda$ scheme. The number of laser field, $n$, required will be determined via the relation $n=\frac{N-1}{2}$ where $N=3,5,7$ is the number of atomic levels. From optical transitions of Fig.~(\ref{EnergyLevel}) it is with the choice of  $^{3}H_{4}\left(\pm\frac{5}{2}\right),^{3}H_{4}\left(\pm\frac{3}{2}\right),^{3}H_{4}\left(\pm\frac{1}{2}\right)$ respectively, as the ground states of the three type ions, with which in turn the five level and seven level system shown in Fig.~(\ref{MLevel}b-c)) are to be realized.

Before proceeding with the description of the Hamiltonian, we point that another experimental setup to which all our calculations are relevant is a series of coupled quantum dots, where the inter-dot tunneling amplitude can be gate-controlled and manipulated adiabatically. Such a manipulation of the tunneling barriers can lead to so-called coherent tunneling adiabatic passage (CTAP), which was suggested as a method for coherent control of electrons in quantum dots \cite{Greentree:2004aa,menchon2016spatial,kral2007colloquium}. As the Hamiltonian is exactly the same as we present below, it follows that the CTAP setup can be an interesting model system to experimentally verify our predictions. 

The Hamiltonian under rotating wave approximation (RWA) and in the interaction picture is given by \cite{Bergmann2001,shore2008}.
\begin{equation}
\begin{aligned}
H \left(t\right)=&\sum_i \delta_{i-1}|i\rangle\langle i|-\left(\sum_i\Omega_i(t) |i\rangle\langle i+1|+h.c\right )
\end{aligned}
\label{Hamit}
\end{equation}
with $|i\rangle\langle i|$ projection matrices, $\Omega_i (t) $ is the (time-dependent) Rabi frequencies for the transitions $i\rightarrow i+1$, and $\delta_{i-1}$ represents $(i-1)$ photon detuning, and $\delta_{0}=0$. We refer the reader to Refs.~\cite{2013arXiv1306.2132G, shore2008} for details about the conditions of adiabaticity. The eigenvalues and the dark-bright states specifically for M-type five level are given in Ref. \cite{2013arXiv1306.2132G} and for general N-level system detailed in \cite{PhysRevA.37.3000}. 

Adiabatic evolution of the states is assured when the rate of non-adiabatic coupling is small compared to the separation of the corresponding eigenvalues, \cite{shore98}. If we define the detuning parameter as the difference between the Rabi frequencies and the corresponding energy level differences, $\Delta_i =|\Omega_i- (\delta_{i+1}-\delta_{i})|$, it follows that for smooth pulse shape with non zero detuning, i.e $\Delta\neq0$, the adiabatic condition is given to be (with the pulse width $\sigma$ and Rabi frequency $\Omega$ where $\Omega=\sqrt{\Omega_p^2+\Omega_s^2}$, with $p=1,3,5$ and $s=2,4,6$)\cite{klein2008experimental}.
\begin{equation}
\begin{aligned}
\big||\Delta|-\sqrt{\Delta^2+\Omega^2}\big|\sigma\gg&1 \label{timeom}
\end{aligned}
\end{equation}
To reiterate, the adiabatic condition given by Eq.\eqref{timeom} expresses that with coherent laser pulses, adiabaticity is maintained by strong coupling (i.e large Rabi frequencies) and smooth temporal laser profiles. With such restriction the state vector of the system closely follows the changes of the adiabatic state. This in turn means that the state vector has to align to specific adiabatic state at all times. In case of STIRAP this adiabatic state is the dark state as will be discussed in the next section (see section \ref{sec:results}). It is also to be noted here that the adiabatic condition for resonance case, i.e ~$\Delta=0$, takes the form $\Omega\sigma\gg1$.

The wave function for  N-level system is a linear combination of its  possible components, with time-dependent coefficients \cite{PhysRevA.37.3000, shore2011manipulating,PhysRevA.66.033405}
\begin{equation}
\begin{aligned}
\Psi \left(t\right)=&\sum_i ^N c_i\left(t\right)|i\rangle, 
\end{aligned}
\label{Psi}
\end{equation}
where $c_i\left(t\right)$ are the probability amplitudes and  they are governed, in the RWA, by the Schr\"{o}dinger time-dependent equation, 
\begin{equation}
\begin{aligned}
\imath\hbar\frac{d }{dt}c\left(t\right)=&  H\left(t\right)c\left(t\right)+i\Gamma c\left(t\right)
\end{aligned}
\label{dcbydt}
\end{equation}
 where $\Gamma$ is the phenomenological dissipation term involving radiative decay process and pure dephasing. Equation\eqref{dcbydt} is solved numerically where parameters of the simulation are given in reduced units. The evolution of population and coherence dynamics is calculated using the probability amplitude $c_j\left(t\right)$ with $|c_j\left(t\right)|^2$ being the probability of finding the system in state $j$ ($\rho_{jj}\left(t\right)$ in the density matrix formalism); while $c_j\left(t\right)c^*_k\left(t\right)$ is the coherence ($\rho_{jk}\left(t\right)$ in the density matrix formalism).  In the calculation experimental parameters such as the life time of the excited state is $164\mu s$, the life time of the two ground states $100s$, and the width of the laser pulse $10\mu s$ have been used. Moreover, the laser pulses have Gaussian profile with the peak laser intensity, $\Omega_0$, being $\Omega_0=\frac{30}{\sigma}$, where $\sigma$ is the width of the laser pulse. 
 
 One can extend the results of five-level M-type and seven-level system into an $N$-level, where we consider an odd number of $N$. In doing so we note that every state is connected to at most two states, except the two end states (initial and target states) which have one link. In this paper we regard one of the end states as state $0$, and the numbering proceeds sequentially along the links to state $N-1$, referred to as the chain link~\cite{shore2011manipulating}. 
 
\section{Population and Coherence Transfer}
\label{sec:results}
\begin{widetext}
\begin{figure*}[htp]

\centering
(a)\includegraphics[width=.35\textwidth]{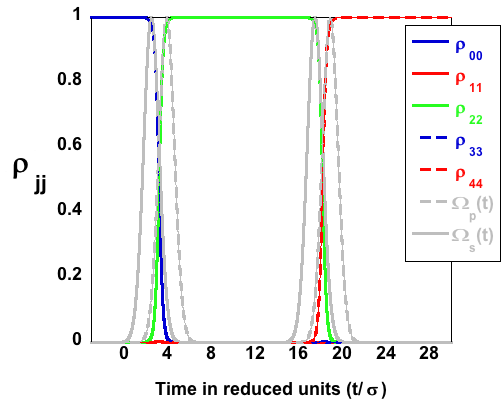}
(b-c)\includegraphics[width=.55\textwidth]{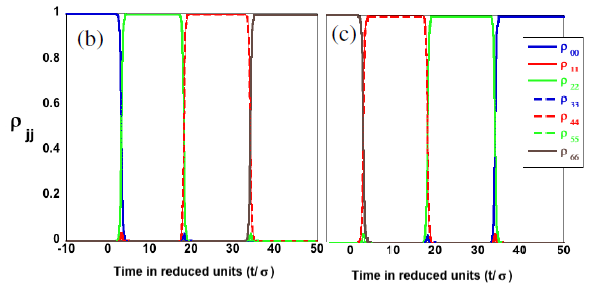}

\caption{(a) Populations in an M-type five level system as a function of time in the presence of a STIRAP protocol. The system is initially prepared in the ground state $|0\rangle$, and then a STIRAP pulse protocol is applied twice (at different times). The first sequence transfers the populations from state $|0\rangle$ to $|2\rangle$, and then the population is transferred to state $|4\rangle$ by applying the second sequence of the protocol. 
(b) Populations in a seven level system. Extending the protocol of (a) with an additional STIRAP pulse, the populations can be transferred throughout the entire system. (c) Populations in seven level system, with a reversed STIRAP protocol: the system is initially prepared in state $|6\rangle$ and the population is transferred to state $|1\rangle$ via successive STIRAP pulses. For all the plots we note that $\rho_{jj}=c_j^*c_j$, the gray lines show the pump (dashed) and Stokes (solid) pulses in arbitrary units.}
\label{fig:Pop57SP}
\end{figure*}
\end{widetext}

We start by introducing the ability to coherently control the transfer of populations and coherences along the multilevel system. This is essential, since the populations and coherences (and their subsequent measurement) are the building blocks of the computation scheme proposed here. For that aim, we introduce a STIRAP protocol, which is an adiabatic process that provides complete coherent transfer in a $\Lambda$-type quantum system. By preparing the system to be initially on the ground state $|0\rangle$, one can  transfer 100\% of the population to state $|2\rangle$ \cite{Vitanov:2001aa}. To do so, the STIRAP protocol applies a counter-intuitive pulse sequence, with the first pulse coupling the unpopulated states $|1\rangle$ and $|2\rangle$ (Stokes pulse), and the second pulse coupling the states  $|0\rangle$ $\& ~|1\rangle$ (pump pulse). An example for the pulse sequence is plotted in Fig. \ref{fig:Pop57SP}a (solid gray line - Stokes pulse, dashed gray line - pump pulse). STIRAP is robust in that it is rather insensitive to the decaying intermediate state $|1\rangle$ \cite{shore2008,Wang2016,Torosov2014,Wang2013}. 

 \emph{Fractional} STIRAP (FSTIRAP) is variant of STIRAP that can be used to create a superposition of states \cite{Vitanov:2001aa, vitanov1999creation}. We aim at creating a coherent superposition between two hyperfine levels, instead of transferring 100\% population between the states. To do so we have to interrupt the STIRAP evolution.  In the FSTIRAP scheme, like that of STIRAP, the Stokes pulse precedes the Pump pulse, but the two pulses vanish smoothly and simultaneously. If, for example, all the population is initially  prepared to be in the ground state, the Stokes pulse is applied when $t\rightarrow-\infty$  but the pump pulse is not yet applied, and therefore  $\frac{\Omega_P\left(-\infty\right)}{\Omega_S\left(-\infty\right)}\rightarrow0$. After a time, i.e when $t\rightarrow+\infty$, we interrupt the pulses in such a manner as to get a constant from their ratio. A convenient realization of the condition  using three pulses, a pump pulse and two Stokes pulses (one with the same time dependence as the pump and another coming earlier) is provided by Vitanov et al. \cite{vitanov1999creation} 

Fig. ~\ref{fig:Pop57SP} exemplifies the STIRAP protocol for a multilevel system. Figures (\ref{fig:Pop57SP}a) and (\ref{fig:Pop57SP}b) show the numerical solution for population transfer as a function of time, for multilevel systems with five and seven states, respectively. The system is initially prepared to be on the state $|0\rangle$, such that the population is transferred from left to right. The gray lines show the pump (dashed) and Stokes (solid) pulses.  
In a similar manner one can initially prepare the system to be on the right most level, for example state $|6\rangle$ in the seven level system, and transfer the population back to state  $|0\rangle$, as shown in Fig (\ref{fig:Pop57SP}c). In such case the coupling order will be as follows: one first couple states $|4\rangle\leftrightarrow|5\rangle$ and then later drive the transition $|5\rangle\leftrightarrow|6\rangle$, and so on.  The ability to coherently manipulate populations is the central tool for generating FF and SISO, as we show in what follows.

\begin{figure}[h!]
	\centering
	\includegraphics[width=3.0 in]{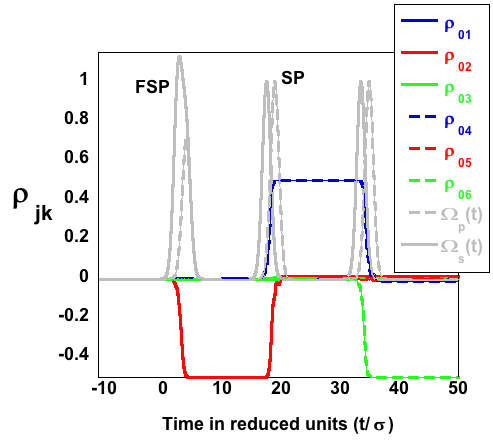}
	\caption{Coherences in a seven level system. The system is initially prepared in the ground state $|0\rangle$ and then the FSTIRAP protocol is applied once, followed by two successive  STIRAP  pulse protocols. The first sequence creates coherence between states $|0\rangle\leftrightarrow|2\rangle$ by distributing populations equally on states $|0\rangle$ and $|2\rangle$. The population of state $|2\rangle$ is then transferred to state $|4\rangle$ by applying a second sequence of the protocol (STIRAP pulse), thereby creating  coherence between states $|0\rangle\leftrightarrow|4\rangle$. The third sequence takes the populations from state $|4\rangle$ to state $|6\rangle$, again allowing the creation of coherences between states $|0\rangle\leftrightarrow|6\rangle$. The plotted coherences are the real part of the coherences ( imaginary parts of the coherences are substantially smaller and are not shown in the plot), the gray lines show the pump (dashed) and Stokes (solid) pulses in arbitrary units.}
	\label{Chrtr7FSPSP}
\end{figure}

By appropriate choice of pulse profile, i.e using FSTIRAP, we are able to distribute the population into different level of quantum states and thereby manage to not only create superpositions of states, but also transfer coherences.  This is demonstrated in Fig (\ref{Chrtr7FSPSP}), in which the coherences are plotted as a function of time for the protocol. The system is initially prepared to be in state $|0\rangle$, and by applying pulse profile of FSTIRAP (gray lines) coherences between the hyperfine levels $|0\rangle\leftrightarrow|2\rangle$ are created. Next, by using the STIRAP protocol we transfer the population from state $|2\rangle$ to $|4\rangle$ (then to $|6\rangle$), directly leading to the \emph{transfer of coherences} between $|0\rangle\leftrightarrow|4\rangle$, and then $|0\rangle\leftrightarrow|6\rangle$. Put differently, once coherence between $|0\rangle$ and $|2\rangle$ is established (using the FSTIRAP pulse protocol), this coherence is now "dragged" along the populations of the higher-placed levels $|4\rangle$ and $|6\rangle$, which are changed using the STIRAP pulse protocol. This prediction of "coherence dragging" can be directly verified experimentally \cite{Beil:2009aa}. 

In what follows, the instantaneous eigenstates of the system, which one finds by diagonalizing the Hamiltonian, were used to describe the population and coherence transfer in the adiabatic basis. This set of adiabatic states includes a dark state $|a_0\rangle$, which contains no contribution from the optically excited state, and the two bright states $|a_+\rangle$ and  $|a_-\rangle$, which are superpositions of all three bare states. Next, bearing in mind that the multilevel system at hand is constructed by concatenating each dopant (with three level $\Lambda$ scheme), the scheme can be viewed to be a step-by-step process of $3\rightarrow5\rightarrow7\rightarrow9\ldots\rightarrow N$ where $N$ is odd. 

For STIRAP formalism pump and Stokes laser pulses are applied in a counterintuitive pulse sequence -- i.e., the Stokes pulse precedes the pump pulse. If, for instance, this protocol is employed in the first step Eqs.\eqref{estate1} and \eqref{angle1} dictates that initially (when $t\rightarrow-\infty$) state $|a^{i}_0\rangle$, where the superscript $i$ indicates first step, is equal to the bare state $|0\rangle$. For all the population is initially prepared to be in this bare state, it in turn means that the system is prepared in the dark state $|a^{i}_0\rangle$. As the evolution is adiabatic, the system remains trapped in this dark state during the interaction. Finally, after the interaction (when $t\rightarrow+\infty$), the dark state $|a^{i}_0\rangle$ is equal to the bare state $|2\rangle$. Therefore the adiabatic state $|a^{i}_0\rangle$, commonly known as dark state, bridges the population transfer from the initial state to the target state. correspondingly, in STIRAP and its derivatives population transfer is realized by following the state vector $|a_0\rangle$. It is also possible to achieve population transfer, with lesser efficiency, following intuitive path via the bright states $|a_\pm\rangle$, this protocol is labeled as \emph{b-STIRAP} in \cite{klein2008experimental}.

So for the first step in this step-wise description-- involving the first three level system, the state vector (dark state) takes the form
\begin{equation}
\begin{aligned}
|a^{i}_0\rangle=&\cos\theta_1 |0\rangle-\sin\theta_1  |2\rangle \label{estate1}\\
\end{aligned}
\end{equation}
where the mixing angle $\theta_1$ is defined to be 
\begin{equation}
\begin{aligned}
\tan\theta_1=&\frac{\Omega_1\left(t\right)}{\Omega_2\left(t\right)},\label{angle1}\\
\end{aligned}
\end{equation}
It is worth pointing out here that the adiabatic state $|a^{i}_0\rangle$ does not involve the optically excited state $|1\rangle$ implying that it is not affected by stimulated emission from this state, one can thus  achieve  complete population transfer  from the ground state $|0\rangle$ to the target state $|2\rangle$. To summarize, in the first step (step-i), the probability amplitude of state $|0\rangle$ is transferred to a linear combination of states $|0\rangle$ and $|2\rangle$. In the second step  (step-ii) the laser pulses $\Omega_1, \Omega_2$ are off while $\Omega_3, \Omega_4$ are on during which the probability amplitude of state $|0\rangle$ is unchanged with the probability amplitude of state $|2\rangle$ is transferred to linear combination of states $|2\rangle$ and $|4\rangle$. Following same argument, for the third step  (step-iii),  the laser pulses $\Omega_1, \Omega_2, \Omega_3, \Omega_4$ are off while $\Omega_5, \Omega_6$ are on during which the probability amplitude of states $|0\rangle, |2\rangle$ are unchanged while the probability amplitude of state $|4\rangle$ is transferred to linear combination of states $|4\rangle$ and $|6\rangle$, thus the state vectors for these two steps are given to be 
\begin{subequations}
\begin{align}
|a^{ii}_0\rangle=&\cos\theta_1 |0\rangle-\sin\theta_1 (\cos\theta_2 |2\rangle-\sin\theta_2  |4\rangle) \label{estate2}\\
\begin{split}
|a^{iii}_0\rangle=&\cos\theta_1 |0\rangle-\sin\theta_1\cos\theta_2 |2\rangle\\
&+\sin\theta_1\sin\theta_2   (\cos\theta_3 |4\rangle-\sin\theta_3  |6\rangle)\label{estate3}
\end{split}
\end{align}
\end{subequations}
where the mixing angles $\theta_2, \theta_3$ are now defined to be 
\begin{subequations}
\begin{align}
\tan\theta_2=&\frac{\Omega_3\left(t\right)}{\Omega_4\left(t\right)},\label{angle2}\\
\tan\theta_3=&\frac{\Omega_5\left(t\right)}{\Omega_6\left(t\right)},\label{angle3}
\end{align}
\end{subequations}
Notice here also that the adiabatic states $|a^{x}_0\rangle, x=ii, iii$ are independent of the lossy states $|3\rangle \& |5\rangle$. 

Now the magnitude of the coherences in the bare states can be obtained using Eqs.\eqref{estate1},\eqref{estate2}, and \eqref{estate3} along with the relation  $|\rho_{jk}|=|c_jc^*_k|$, where $c_k=\langle k|a^x_0\rangle$  and $x=i,ii,iii$. Hence the amplitudes of the coherence created, i.e $|\rho_{jk}|$,  can be calculated in terms of the time dependent mixing angles at each steps -- i.e $\theta_1, \theta_2, \& ~\theta_3$, doing so yields 
\begin{subequations}
\begin{align}
|\rho_{02}|=&|\cos\theta_1\sin\theta_1\cos\theta_2|\label{r02}\\
|\rho_{04}|=&|\cos\theta_1\sin\theta_1\sin\theta_2\cos\theta_3|\label{r04}\\
|\rho_{06}|=&|\cos\theta_1\sin\theta_1\sin\theta_2\sin\theta_3|\label{r06}\\
|\rho_{24}|=&|\sin\theta_1\cos\theta_2\sin\theta_1\sin\theta_2\cos\theta_3|\label{r24}\\
|\rho_{46}|=&|\sin^2\theta_1\sin^2\theta_2\cos\theta_3\sin\theta_3|\label{r46}\\
|\rho_{26}|=&|\sin^2\theta_1\cos\theta_2\sin\theta_2\sin\theta_3|\label{r26}
\end{align}
\end{subequations} 
Moreover the mixing angles given in Eqs.\eqref{angle1}, \eqref{angle2}, and \eqref{angle3} varies at each step depending on the type of pulse used
\begin{equation}
\begin{aligned}
\hline 
\textbf{Step-i} && \textbf{Step-ii} && \textbf{Step-iii}\\
\hline 
\\
\theta_1=&\frac{\pi}{4} & \theta_1=&\frac{\pi}{4} & \theta_1=&\frac{\pi}{4} \\
\theta_2=&0 & \theta_2=&\frac{\pi}{2} & \theta_2=&\frac{\pi}{2} \\
\theta_3=&0 & \theta_3=&0 & \theta_3=&\frac{\pi}{2} \\
\hline
\end{aligned}
\label{stepstheta}
\end{equation}

Therefore from Eqs.\eqref{r02}-\eqref{r26} and the value of mixing angles at each step, Eq.\eqref{stepstheta}, the coherences for seven level system is seen to be transferred as $\rho_{02}\rightarrow\rho_{04}\rightarrow\rho_{06}$ as shown in Fig (\ref{Chrtr7FSPSP}). 

Note that initially $\cos\theta_1=\sin\theta_1=\frac{1}{\sqrt{2}}$ and $\sin\theta_2=\sin\theta_3=0$ consequently only $\rho_{02}\equiv\frac{1}{2}$ survives from Eqs.\eqref{r02}-\eqref{r26}. At the next step while $\theta_1, \theta_3$ remains same as step-i, but $\theta_2$ changes leading to $\cos\theta_2=0$, and  this means now $\rho_{02}$ vanishes and $\rho_{04}\equiv\frac{1}{2}$ is the surviving coherence. Lastly, at the end of the interaction, $\theta_1,\theta_2$ keeps same value as step-ii, yet $\theta_3=\frac{\pi}{2}$,  at this stage making use of $\sin\theta_2=\sin\theta_3=1$, one readily see that $\rho_{06}\equiv\frac{1}{2}$ is the sole surviving coherence, from which follows that the coherences for seven level system is seen to be transferred as $\rho_{02}\rightarrow\rho_{04}\rightarrow\rho_{06}$

\section{Realization of logic machines}
\label{sec:logicmachine}
Flip-flops allow sequential circuits to have a long-lived, readable state (i.e., memory), which is something that combinational logic circuits do not have, and therefore require a different design than logic elements. In this section, the design of FFs using the multilevel systems and their manipulation described above is introduced.
\subsection*{Toggle  Flip Flop}
\label{ssec:TFF}
Generally, a flip flop has two inputs, where one of the inputs is a control input, and two outputs. For a Toggle FF (TFF) the control input is labeled $T$, and the output is the bit that is stored (labeled $Q$). The characteristic table for a TFF is given in Table~(\ref{table:Tflipflop}).
\begin{table}
\centering
\begin{tabular}{ |l| l| l | |l|  }
\hline
 \textbf{T} & \textbf{$Q$} & \textbf{$Q(t+1)$} &  \textbf{Remark}\\
\hline
0  & 0 & 0 & Hold \\
0  & 1 &  1 & Hold\\
1  & 0 &  1 & Toggle  \\
1  & 1 & 0 &  Toggle  \\
\hline
\textbf{Pulse} & $s(t)$ & \textbf{$s(t+1)$} & \\
\hline
\end{tabular}
\caption{Characteristic table for T-flip flop, where $s(t)$ and $s(t+1)$ stand for the current state and the next state, respectively. }
\label{table:Tflipflop}
\end{table}
The TFF has two possible values, $T = 0$ or $T = 1$. When $T = 0$, the flip flop performs a "hold", meaning that the output, $Q$ is kept the same as it was before the operation (STIRAP pulse in our case). When $T = 1$, the flip flop performs a "toggle", meaning that the output $Q$ is negated after the pulse, compared to the value before the pulse. Hence, in a TFF, one can either maintain the current state's value for another cycle, or one can toggle the value (i.e negate it) after pulse is applied. 

To implement the TFF we consider the state of the three level $\Lambda$-system, before and after a pulse is applied, as present state and next state of the FF respectively, and define the pulse as a control input. The logic assignment for pulse is \textbf{0} if the pulse is \textbf{OFF} and logic value \textbf{1} if the pulse is \textbf{ON}. For the present and next states the following logic assignment are made: if all the populations are in state $|0\rangle$ the corresponding logic value is \textbf{0}, and if it is in state $|2\rangle$ the corresponding logic assignment is \textbf{1}. Recall that we use the STIRAP mechanism to transfer populations between states $|0\rangle$ and $|2\rangle$. To this end we refer the pulse sequence of the STIRAP protocol as \textbf{PULSE} in the following. If the population is prepared to be initially in state $|0\rangle$ it remains in state $|0\rangle$, i.e it holds, if we do not apply \textbf{PULSE}, but will be transferred to state $|2\rangle$, i.e toggle,  if we apply \textbf{PULSE}. This fulfills the first and third rows of the characteristic table in Table~(\ref{table:Tflipflop}). If the state is initially prepared to be  $|2\rangle$ it remains in state  $|2\rangle$, if we do not apply \textbf{PULSE}, and will be transferred, i.e toggle, to state  $|0\rangle$ if \textbf{PULSE} is applied, providing the second and forth row of the characteristics table.

In the realization of a TFF presented here only one output storage is shown, because the output of a FF is just the bit that is stored, i.e $Q$, and the second  output is just the negation of the stored bit information, i.e $\overline Q$, and therefore the second output do not need a separate storage. 
The TFF characteristic table, Table~(\ref{table:Tflipflop}),  resembles the truth table for a \textbf{XOR} gate, where the output is \textbf{1} if and only if one of the inputs are \textbf{1}, but not both. We refrain from calling the characteristic table a \textbf{XOR} for the following reasons. First, \textbf{XOR} has two inputs, and one output, but a TFF has a control input and a control output. Second, note that the second column and the third column are really the same output, but at different points in \emph{time}, a requirement of the FF.  
\subsection*{Delay  Flip Flop}
\label{ssec:DFF}
The Delay flip flop (DFF) behaves like a delay element; the next state of this device is equal to its present excitation. It is therefore characterised by \cite{kohavi2010switching}
\begin{equation}
\begin{aligned}
Q(t+1)=&D(t)~~.
\end{aligned}
\label{DFFeqn}
\end{equation}
Equation~\eqref{DFFeqn} basically states that data present when the pulse is applied is transferred from $D$ to $Q$. Just like TFF, the DFF also has two possible values, \textbf{0} and \textbf{1}. We can see from the characteristic table in Table~(\ref{table:Dflipflop}) that when $D=0$, the flip flop resets, meaning the output $Q$ is set to \textbf{0}. But, if $D=1$, the flip flop sets, meaning the output is set to \textbf{1}.
\begin{table}
\centering
\begin{tabular}{ |l| l| l| |l | l| }
\hline
&  &  &  & \\
 \textbf{D} & $Q$ & $Q(t+1)$ & $\overline{Q(t+1)}$ & \textbf{Remark}\\
\hline
0  & 0 & 0 & 1 & Reset \\
0  & 1 &  0 & 1 & Reset\\
1  & 0 &  1 & 0 & Set  \\
1  & 1 & 1 &  0 & Set  \\
\hline
\textbf{Pulse} & $s(t)$ & \textbf{$s(t+1)$} &  &\\
\hline
 & Initial & coherence & population  & \\
 & state & between  & in $|0\rangle$ or & \\
 &   & $|0\rangle\leftrightarrow|2\rangle$ &  population  & \\
  &   &  &  in $|2\rangle$  & \\
\hline
\end{tabular}
\caption{Characteristic table for D-flip flop, where $s(t)$ and $s(t+1)$ stand for the current state and the next state, respectively. }
\label{table:Dflipflop}
\end{table}

Let us describe how the three level $\Lambda$ system can define a DFF. The control input is the applied pulse, which in this case is the STIRAP sequence and its variant FSTIRAP . The pulse is assigned to be the control input, \textbf{(D)}, with logic assignment of \textbf{0} if the \textbf{STIRAP} pulse profile is used and \textbf{1} if \textbf{FSTIRAP} pulse profile is used. The initial state of the system can be prepared to be either in state $|0\rangle$ or $|2\rangle$. The logic value \textbf{0} is assigned if the system is initially prepared to be in state $|0\rangle$ and \textbf{1} if it is prepared in state $|2\rangle$. This initial state corresponds to the present state $Q(t)$ of the DFF. The next state, which is the output after the pulse is applied, is encoded to be the creation of coherence between states $|0\rangle$ and $|2\rangle$; if coherence is created we assign logic value \textbf{1}, else \textbf{0}. 
\begin{figure}[h!]
\centering
\includegraphics[width=2.50 in]{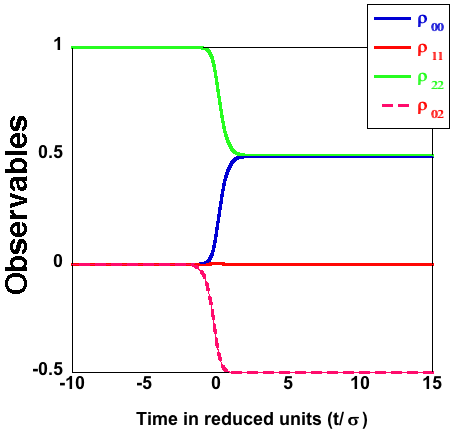}
\caption{System observables, i.e. populations and coherences, during the FSTIRAP pulse profile. The system is initially prepared in state $|2\rangle$.  Then, the states $|0\rangle\leftrightarrow|1\rangle$ are coupled by a pump pulse, and later the states $|1\rangle\leftrightarrow|2\rangle$ are coupled by a Stokes pulse. By applying the FSTIRAP protocol, coherences between states $|0\rangle$ and $|2\rangle$ are created. Blue, red, green, and dashed red respectively represent  $\rho_{00}, \rho_{11}, \rho_{22}$, and $\Re\rho_{02}$}
\label{ObservFSP2}
\end{figure}

Coherence between states $|0\rangle$ and $|2\rangle$  is created if a FSTIRAP pulse profile is applied, as shown in Fig.(\ref{ObservFSP2}).  On the other hand, if a STIRAP pulse is applied no coherence is created, since the protocol transfers the population completely from the initial state to target state.  To create coherence between state $|0\rangle$ and $|2\rangle$ the populations need to be distributed between these states. Therefore, if the system is prepared to be in either of the states $|0\rangle$ and $|2\rangle$ and STIRAP is applied, no coherence will be generated and hence the output for both cases is \textbf{0}. This condition satisfies the first and second rows of Table~(\ref{table:Dflipflop}). If the FSTIRAP pulse profile is applied when the system is  initially prepared to be in either of the states $|0\rangle$ and $|2\rangle$, then coherence between states $|0\rangle\leftrightarrow|2\rangle$ is generated, corresponding to the third and fourth rows of Table~(\ref{table:Dflipflop}). 

The fourth column in Table~(\ref{table:Dflipflop}) is the negation of the output, i.e. $\overline{Q(t+1)}$. For the above case, it corresponds to the populations in either of states $|0\rangle$ or $|2\rangle$. When the STIRAP pulse profile is applied, the populations are transfered completely, and the output will have logic value \textbf{1}. When the pulse profile of FSTIRAP is used, it distributes the populations equally between the states $|0\rangle$ or $|2\rangle$. In this case, we assign the corresponding logic value to be \textbf{0}. Simply put: if the population is localized in either $|0\rangle$ or $|2\rangle$, the corresponding output is logic value \textbf{1}, else it is \textbf{0}. This will generate the second output of the DFF (at the next unit of time), i.e. $\overline{Q(t+1)}$. It is worth stressing here that the fact that we are using coherences for a finite-state machine and not only population is quite appealing, as it increases the number of observables that can be used for input-output variables. Having more observables, consequently, is advantageous if one wants to perform parallel logic, because one can make use of each of the observables to perform logic operations in parallel as has been demonstrated in~\cite{Fresch:2013aa}.
\subsection*{Serial-in Serial-out Shift registers}
\label{ssec:SISO}
\begin{figure}[h!]
\centering
\includegraphics[width=2.50 in]{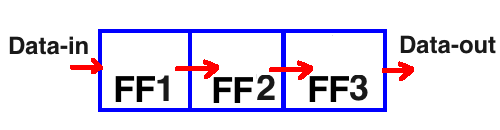}
\caption{Schematic representation of SISO shift (right) register, where FF1, FF2, FF3 represent flip flops 1, 2, and 3 respectively. In such scheme input data will be propagated to the right .}
\label{SISO}
\end{figure}

A shift register essentially comprises of multiple single bit flip flops (FF), one for each data bit connected together in a serial chain arrangement so that the output from one FF becomes the input of the next FF and so on, resulting in a circuit that shifts by one position (with each signal of the clock) the bit data stored in it. A Serial-in Serial-out (SISO) shift register is the simplest kind of shift registers. In such shift registers the data string is presented at 'Data in', see Fig.(\ref{SISO}), and is shifted to the right after each clock (in digital circuit clock refers to a periodic signal alternating between $0$ and $1$). 

A basic three-bit shift register can be constructed using three DFFs, as shown schematically in Fig.(\ref{SISO}).  The operation of the circuit is as follows.  The register is first cleared, forcing all three outputs to zero.  The input data is then applied sequentially to the $D$ input of the first flip-flop on the left (FF1).  During each clock pulse, one bit is transmitted from left to right.  The least significant bit of the data has to be shifted through the register from FF1 to FF3. We propose the implementation of such shift registers by concatenating the three level $\Lambda$ type system to form an $N$-level system. We focus our discussion in this paper only to five-level and seven-level systems, Figs.(\ref{MLevel}b), (\ref{MLevel}c), but the proposed design can be extended to $N$-level systems, where $N$ is odd. 

\begin{figure}[h!]
\centering
\includegraphics[width=2.5 in]{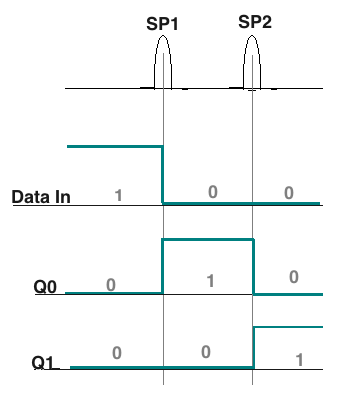}
\caption{Schematic representation of the output  $Q_0$ and $Q_1$ given Data In$=100$. As sequence of two pulses is applied input data is seen shifting to right as $100\rightarrow010\rightarrow001$}
\label{DtoQ5}
\end{figure}

The design and implementation of SISO shift registers using the $N$-level system is as follows. The "Data In" for the shift registers is encoded as the initial state of the system. If , for instance,  a five-level system shown in Fig.(\ref{MLevel}b) is considered, we form the "data word" by reading the population at levels $|0\rangle, |2\rangle$, and $|4\rangle$ and place it as a string with the order $|0\rangle,|2\rangle,|4\rangle$. We further impose the constraint, that populations initially reside in either of  state $|0\rangle, |2\rangle$, or $|4\rangle$. The logic value \textbf{1} is assigned if all the population are in one of the states only, else it is \textbf{0}. Therefore, although our three bit string can have eight possible combinations, the constraint excludes the strings $011,101, 110, 111$, since two (or more) states cannot simultaneously have 100\% population. Hence, we are left with the following "data words" to consider for "Data In", $000, 001,010, 100$. If for example we initially prepare the system to be on state $|0\rangle$ the corresponding "Data in" is $100$ (see Fig.(\ref{DtoQ5})), set by the order to be populations in $|0\rangle,|2\rangle,|4\rangle$. Similarly, the strings $001,010$ correspond to populations initially prepared in $|4\rangle$ and $|2\rangle$ respectively. 

Next, the clock pulse of the shift register is encoded by the pulse sequence we apply. Because we demand a complete population transfer between states $|0\rangle, |2\rangle$, or $|4\rangle$ we readily use the pulse protocol of  STIRAP to achieve this goal. In doing so we apply STIRAP pulse sequence (here after \textbf{SP}) twice. If we initially prepare the system to be in state $|0\rangle$ the first pulse sequence, let us label it \textbf{SP1} (where the \textbf{P} couples states $|0\rangle\leftrightarrow|1\rangle$ and the \textbf{S} couples states $|1\rangle\leftrightarrow|2\rangle$) transfers the populations to state $|2\rangle$. The second pulse sequence, which is labeled \textbf{SP2} (where the \textbf{P} now couples states $|2\rangle\leftrightarrow|3\rangle$ and the \textbf{S} couples states $|3\rangle\leftrightarrow|4\rangle$) transfers the populations to state $|4\rangle$. Therefore we see the population has shifted following the state path $|0\rangle\rightarrow|2\rangle\rightarrow|4\rangle$ as shown in Fig. (\ref{fig:Pop57SP}a). 
\begin{figure}[h!]
\centering
\includegraphics[width=3.5 in]{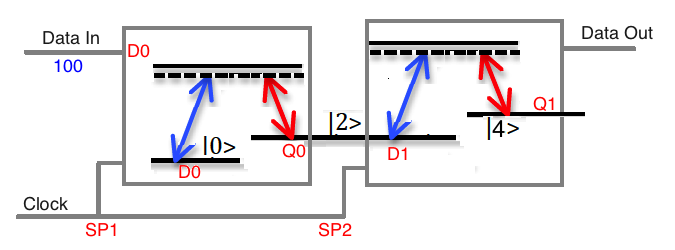}
\caption{Schematic representation of the logic assignment and implementation of two cascaded flip-flop SISO shift (right) register using a five level system (see text). }
\label{SISO5}
\end{figure}

To elaborate more, let us correlate the physical system at hand and the FFs as shown in Fig.(\ref{SISO5}). The system is initially prepared to be in state $|0\rangle$, this means the "Data In" is encoded to be the population in state $|0\rangle$. In FFs the string of "Data In" is transferred to $D_0$, meaning it is delayed at the first FF. We encode this delay element $D_0$ to be again the population of state $|0\rangle$ at time $t$.  After applying the first pulse sequence, \textbf{SP1} (at time $t+1$), the population is transferred to state $|2\rangle$, see Fig. (\ref{fig:Pop57SP}a). The population of state $|2\rangle$ (at time $t+1$) is encoded to be the output $Q_0$. For the second stage of the operation, i.e the second FF, state $|2\rangle$ (at time $t+1$) serves as an initial state and thus is encoded to be $D_1$. Next the second sequence of the pulse , \textbf{SP2} (at time $t+2$), is applied consequently transferring all the populations to state $|4\rangle$, as seen in Fig. (\ref{fig:Pop57SP}a). Hence the population of state $|4\rangle$ (at time $t+2$) is encoded to be the output $Q_1$ for the FF.

The extension to seven-level system is straight forward with additional output bit encoded to be the population in state $|6\rangle$. It is worth mentioning here that so far the SISO shift registers presented propagate the data towards the right, but it is possible to design the left propagating SISO shift registers by preparing the system to be initially in the state which is at the right hand side , and from there transfer the population to the leftmost state, and thereby shift data to the left.   

One of the interesting features of the setup introduced here is the possibility to use coherences for the design of computing element. As a demonstration, we show the implementation of SISO shift register using coherences in seven-level system. The system is initially prepared to be in state $|0\rangle$, which serves as the "Data In". It is then is transferred to  $D_0$, where $D_0$ is encoded to be the population of state $|0\rangle$ at initial time $t$. After the first FSTIRAP pulse sequence \textbf{FSP1} (at time $t+1$), half of the population is transferred to state $|2\rangle$, creating coherence between states $|0\rangle\leftrightarrow|2\rangle$ as shown in Fig.(\ref{Chrtr7FSPSP}). The output $Q_0$ is encoded to be the coherence created, i.e $\rho_{02}$, at time $t+1$. $Q_0$ then serves as the input for the second flip-flop and hence $D_1$ is again the coherence created, i.e $\rho_{02}$, at time $t+1$. We now apply a (regular) STIRAP pulse sequence \textbf{SP2} after which, say at time $t+2$,   all the population of state $|2\rangle$ is transferred to state $|4\rangle$. As described in Fig.~\ref{ObservFSP2} applying the STIRAP pulse "drags" the coherence along, yielding an output $Q_1$ (encoded to be the coherence created, i.e $\rho_{04}$ at time $t+2$). Next, a third pulse sequence, i.e second STIRAP pulse \textbf{SP3}, is applied, after which, at time $t+3$,  all the population of state $|4\rangle$ is transferred to state $|6\rangle$. Again, the coherence is dragged along, yielding output $Q_2$ (encoded to be the coherence created, i.e $\rho_{06}$ at time $t+3$). Therefore, the "data word" 0001 is obtained as an output, encoded to be the string $P_{00}P_{02}P_{04}P_{06}$, where $P_{00}$ is the initial population on state $|0\rangle$, and $P_{0k}$ are real parts of the coherences between state $|0\rangle\leftrightarrow|k\rangle$, for $k=2,4,6$. 

Use of the coherences in this particular case has an advantage in that it will enable us access  additional state. In the five level system, for instance, state $011$ in the ordering introduced earlier (i.e $|0\rangle,|2\rangle,|4\rangle$)  is now accessible. In such case coherence transfer is seen to be $\rho_{24}\rightarrow\rho_{04}\rightarrow\rho_{02}$. This means initially the system is prepared to be equally distributed amongst states $|2\rangle,|4\rangle$, and consequently we have an initial coherence created between these states, $\rho_{24}$. By applying the first pulse sequence \textbf{SP1}, which, in this case, first couples empty states $|0\rangle$ and $|1\rangle$ using pump pulse and later apply the Stoke pulse that couples states $|1\rangle$ and $|2\rangle$, with this protocol one is able to transfer the population of state $|2\rangle$ to $|0\rangle$ thereby creating coherences between states $|0\rangle\leftrightarrow|4\rangle$. After which now apply the second sequence of the pulse sequence \textbf{SP2} where now  the pump pulse that couples states $|2\rangle$ and $|3\rangle$  is applied before the stokes pulse that couples states $|3\rangle$ and $|4\rangle$, doing so enables to transfer the population from state $|4\rangle$ to $|2\rangle$ thereby creating coherence between states $|0\rangle\leftrightarrow|2\rangle$

\section{Conclusion}
\label{sec:conclusion}
In conclusion, we have presented an all-optical implementation of TFF and DFF using a three-, five- and seven-level $\Lambda$ systems, and shown the design of a SISO shift register using two and three communicating three-level $\Lambda$ type systems via adiabatic population transfer scheme. Two optical protocols, namely STIRAP and FSTIRAP, were combined to create and transfer populations and coherences across the states of the system for logic operations. In a similar manner, one can obtain multiple cascaded FFs from $N$-multilevel system, where $N$ is odd, by transferring population from one end of the chain to the other. The shift registers in general are bidirectional, meaning that the system can be initially prepared in the left hand end and transfer the population to the right hand end, or vice versa. 

Furthermore, we have presented the use of coherences for the design of atomic-scale memory logic, specifically for designing delay flip-flops and SISO shift registers. The ability to use coherences for logic opens an opportunity for implementation of parallel logic as the inclusion of coherences enlarges the accessible state number for input/output variables. The potential of using coherences for parallel logic operations will be explored in future studies. 

However, care must be taken in making use of coherences for logic operation as the information encoded in them will be washed away by the interaction with the environment. In the case presented in this paper it is worthwhile remembering the logic operations proposed here are memory machines. This means that as long as we have a long-lived coherence in comparison to the computation time of the logic machine, as is the case here, it is possible to make use of coherence serve as memory storing machine. What should be done thus is store and read the information encoded before it is completely lost, and this is viable with fast computing times coupled with $\approx\mu s$ life of the coherences as the proposed scheme.




\bibliography{mybibfile}

\end{document}